\documentclass[doublecol,figures]{epl2}
\renewcommand{\revision}[1]{{#1}}%
\usepackage{amsmath,amssymb}
\usepackage{bm}

\title{Anomalous diffusion analysis of the lifting events in the event-chain Monte Carlo for the classical $XY$ models}
\shorttitle{Analysis of the lifting events in the event-chain Monte Carlo}
\author{Kenji Kimura\inst{1, 2}\thanks{E-mail: \email{kimura.kenji.6u@kyoto-u.ac.jp}} \and Saburo Higuchi\inst{1}\thanks{E-mail: \email{hig@math.ryukoku.ac.jp}}}
\shortauthor{K. Kimura \and S. Higuchi}
\institute{
 \inst{1} Department of Applied Mathematics and Informatics,  Ryukoku University - Otsu, Shiga 520-2194, Japan\\
 \inst{2} Department of Biosystems Science, Institute for Frontier Life and Medical Sciences,  Kyoto University - Kyoto 606-8507, Japan
}

\pacs{02.70.Tt}{Justifications or modifications of Monte Carlo methods}
\pacs{05.40.Fb}{Random walks and Levy flights}
\pacs{75.10.Hk}{Classical spin models}

\abstract{%
We introduce a novel random walk model that emerges in the event-chain Monte Carlo (ECMC) of spin systems.
In the ECMC, the lifting variable specifying the spin to be updated changes its value to one of its interacting neighbor spins. 
This movement can be regarded as a random walk in a random environment with a feedback.
We investigate this random walk numerically in the case of the classical $XY$ model in $1,2$, and $3$ dimensions to find that it is \revision{superdiffusive} near the critical point of the underlying spin system.
It is suggested that the performance improvement of the ECMC is related \revision{to} this anomalous behavior. }

\begin{document}
\maketitle

\section{Introduction}
The Monte Carlo algorithm (MC) is \revision{well-used} method in statistical mechanics.
The most well-known and important class is the local Metropolis MC (LMC)~\cite{metropolis1953calculation} which keeps the detailed balance condition.
In the LMC, however, there are well-known difficulties, for example, the critical slowing down near the critical point.
Therefore it is important to go beyond the detailed balance for improvement of the MCs.

Recently MCs with the broken detailed balance condition are proposed. The examples include event-chain Monte Carlo algorithm (ECMC)~\cite{bernard2009event}, geometric allocation algorithm\cite{suwa2010markov}, and skew detailed balance algorithm~\cite{turitsyn2011irreversible, sakai2013dynamics}.
They are very interesting studies that show improvement \revision{in} the efficiency of the MC. We will analyze the ECMC as a stochastic process in this letter.

The ECMC was initially introduced for hard sphere systems and then was
extended to classical continuous spin systems~\cite{michel2015event, nishikawa2015event, nishikawa2016event}.
It
consists of the factorized Metropolis filter, an additional \revision{degree} of freedom called the lifting variable, and rejection-free algorithm~\cite{peters2012rejection}.
It does not satisfy the detailed balance condition but satisfies the global balance condition~\cite{michel2014generalized}.

In Ref.~\cite{michel2015event}, the ECMC was applied to the $XY$ model and it was shown that the ECMC relaxes more rapidly than the LMC.
It is desirable to understand whether and how the ECMC is efficient for various systems. This motivated us to study the stochastic dynamics of this algorithm as a first step.

In this letter, we will define the \textsl{lifting variable random walk} as that of the movement of the lifting variable that specifies the site to be updated. The lifting variable hops to another site with a probability depending on \revision{the} spin variables which interacts with the spin at the current site.

We believe that studying this walk will help one to understand the dynamics of the ECMC. Moreover, this walk is interesting and is worth studying in its own right. It is a novel random walk in a random environment with a feedback \cite{schulz1999random,schulz2000random}.
We will investigate this random walk numerically in the case of the classical $XY$ model to find that it is \revision{superdiffusive} near the critical point.

\section{The event-chain Monte Carlo}
\label{sec:ecmc}
The ECMC is rejection-free because at rejection the update of an alternative variable is automatically accepted (an `event') in such a way that the global balance is kept~\cite{michel2015event}.
First\revision{,} we consider a spin system and the update of a configuration $a$ to $b$ by the LMC.
The acceptance probability for the Metropolis filter is defined by
\begin{equation}
	p^\text{Met} (a \to b) = \min \left[1, \exp(-\beta \Delta E) \right],
\label{met1}
\end{equation}
where $\Delta E = E^b - E^a$ is the energy change and $\beta$ is the inverse temperature.
In the continuous spin systems with a pairwise interaction, we can transform eq.~\eqref{met1} into
\begin{equation}
	p^\text{Met} (a \to b) = \min \left[1, \prod _{\langle i, j \rangle} \exp \left( -\beta \Delta E_{ij}\right) \right],
\label{met2}
\end{equation}
where $\Delta E_{ij} = E_{ij}^b - E_{ij}^a$ is the pair energy change. 
The update process satisfies the detailed balance condition
\begin{equation}
	\pi_a p^\text{Met}( a \to b) = \pi_b p^\text{Met}(b \to a),
\label{dbc}
\end{equation}
where $\pi$ is the Boltzmann weight $\pi_* = \exp(-\beta E_*)$.

In the ECMC, we design the update process where a spin variable receives persistent infinitesimal \revision{updates}.
Because this violates the detailed balance condition,
we seek to recover the global balance condition. 
We employ the factorized Metropolis filter~\cite{michel2014generalized}
\begin{equation}
	p^\text{fact} ( a \to b) = \prod _{\langle i, j \rangle} \min [ 1, \exp (-\beta \Delta E_{ij}) ].
\end{equation}
It can factorize all individual pair energies.
The physical configuration $a$ and $b$ are the extended to include \textsl{lifting variable} $k$.
It represents the spin currently rotated.
Under \revision{the} factorized Metropolis filter and infinitesimal rotations, the rejection is judged for each interacting spin $\ell$ independently and the first rejection pair $(k,\ell)$ is determined uniquely.
When the first rejection occurs, we change the value of the lifting variable from $k$ to $\ell$. 
Then we say that a lifting event occurs.

For concreteness, we describe the ECMC method for the $XY$ model in detail.
The classical $XY$ model is one of the simplest continuous spin models in statistical mechanics, which is defined by the energy function
\begin{equation}
E^a = 
\sum_{\langle i,j \rangle} E_{ij}=
-J \sum _{\langle i,j \rangle} \bm{s}_i \cdot \bm{s}_j = -J \sum _{\langle i,j \rangle} \cos (\theta_i - \theta _j),
\end{equation}
where $E_{ij}$ is the pair energy, $J$ is the coupling constant, $\bm{s}_*$ is the two-component unit vector, and $\theta _*$ is the rotation angle of $\bm{s}_*$. 
The notation $\langle i,j\rangle$ means all \revision{the} pairs of nearest-neighbor spins.
In two\revision{-}dimensional square lattice case, this model has the Kosterlitz-Thouless transition \cite{kosterlitz1973} at  $\beta = 1.11996(6)$~\cite{komura2012large}.
In three\revision{-}dimensional cubic case, this model has the second order phase transition at $\beta = 0.454166$~\cite{komura2014cuda}.

For infinitesimal rotation \revision{toward} the event angle $\theta _{k, \text{event}}$, we introduce the event-driven approach \cite{peters2012rejection, michel2014generalized}.
In order to determine \revision{$\theta _{k, \text{event}}$}, it is necessary to calculate the increase of the pair energy $\Delta E_{k}(\ell)$ of each pair $(k,\ell)$
\begin{equation}
	\Delta E_{k}(\ell) = -\frac{1}{\beta}\log \gamma_{k\ell},
\end{equation} 
where $\gamma_{k\ell}$ means \revision{a} random number uniformly distributed between $0$ and $1$.
The increase and the event angle $\theta _{k, \text{event}}$ is related by
\begin{equation}
	\Delta E_{k}(\ell) = \int ^{\theta_{k, \text{event}}} _{\theta_{k, \text{current}}} \max \left( 0, \frac{d E_{k\ell}}{d \theta_k} \right) d \theta _k.
	\label{e_int}
\end{equation}
We solve eq.~\eqref{e_int} to determine the event angle $\theta _{k, \text{event}}$ for each $\ell$ interacting with $k$.
Then \revision{we} choose the $\ell$ which gives the minimal angle.
For more detail, see \revision{Refs.}~\cite{nishikawa2015event,nishikawa2016event}.

\section{Lifting variable random walk}
\begin{figure}
\onefigure[width=1\linewidth]{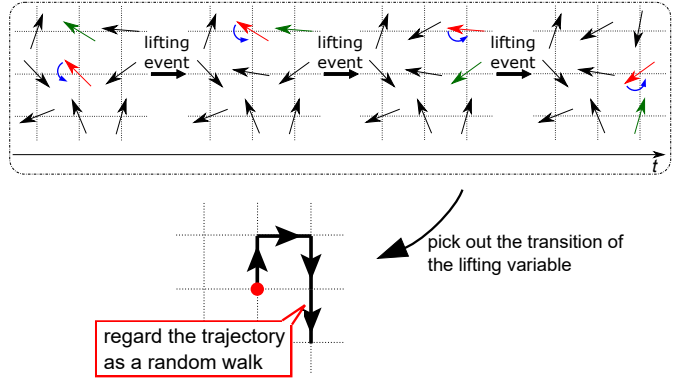}
\caption{\revision{A} brief description of our viewpoint.  Arrows mean the current spin configuration. \revision{The} red arrow is the updated spin specified by the lifting variable. \revision{The} green arrow is the first spin causing the rejection (lifting event).}
\label{lrw}
\end{figure}

In the ECMC of a general spin system with pairwise interactions on a regular or random lattice, the lifting variable $k$ changes its value to one of the spins $\{\ell\}$ interacting with $k$. 
Thus a movement of the lifting variable can be regarded as a random walk (see fig.~\ref{lrw}). 
Its \revision{transition} probability depends on $\bm{s}_k$ and $\bm{s}_\ell$'s. 
We call this random walk \emph{lifting variable random walk}.

This random walk is regarded as a special case of the random walk in a random environment\cite{0305-4470-39-40-R01}. Namely, the environment consists of the spin variables which are not independent each other.
Moreover, it is a random walk in the random environment with \emph{feedback} \cite{schulz1999random,schulz2000random} because the spin at $k$ changes its value depending on the movement of the lifting variable.

\revision{
For the comparison to the above random walk \textsl{with feedback},
we formally define a random walk in a random environment \emph{without feedback}  to the environment spins.
In the walk without feedback, given a spin configuration, the lifting variable shall stochastically move to a neighbor site according to exactly the same rule as the walk with feedback. After that, the current spin $\theta_k$ shall \emph{not} be updated. Therefore it is
a random walk in a quenched random environment and is useless for the measurement of thermodynamic quantities of the spin system. It turns out, however, to be userful for the understanding of the walk with feedback.
}

There are many possible choices of the time unit for the random walk.
In our analysis, time $t$ is defined to be  the number of lifting events. Therefore at every \revision{$t$,} the walker moves to another site and never stays at the same site.

A random walk can be characterized by \revision{the} anomalous diffusion exponent $d_\mathrm{w}$ of the expectation value of mean square displacement (MSD)
\begin{equation}
	\langle X(t)^2 \rangle \revision{= D} t^{2/d_\mathrm{w}},
\label{msd}
\end{equation}
where $X(t)$ means displacement from $X(0)=0$, \revision{$D$ is the diffusion coefficient} and $t$ is \revision{the} time.
We classify diffusion into three classes with the value of $d_\mathrm{w}$: \revision{superdiffusion} $(d_\mathrm{w} < 2)$, normal diffusion $(d_\mathrm{w} = 2)$, and \revision{subdiffusion} $(d_\mathrm{w} > 2)$~\cite{ben2000diffusion, METZLER20001,metzler2004restaurant}. The Brownian motion is classified to normal.
Super and \revision{subdiffusion} are called anomalous diffusion.

\section{Results}
\label{sec:result}
\begin{figure*}
\begin{minipage}{0.325\hsize}
\begin{center}
   \includegraphics[width=1\linewidth]{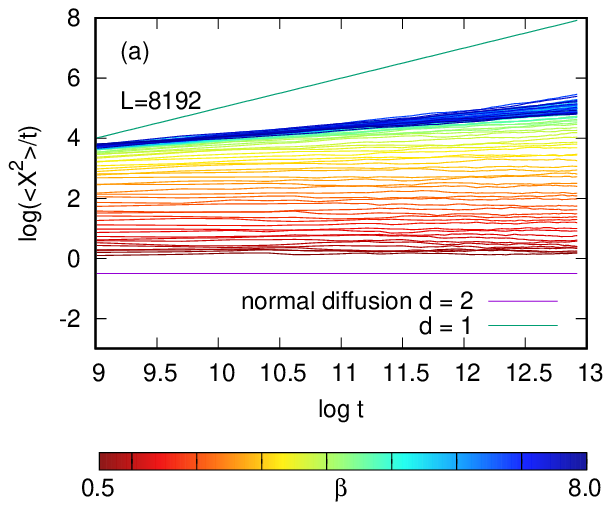}
\end{center}
\end{minipage}
\begin{minipage}{0.325\hsize}
\begin{center}
   \includegraphics[width=1\linewidth]{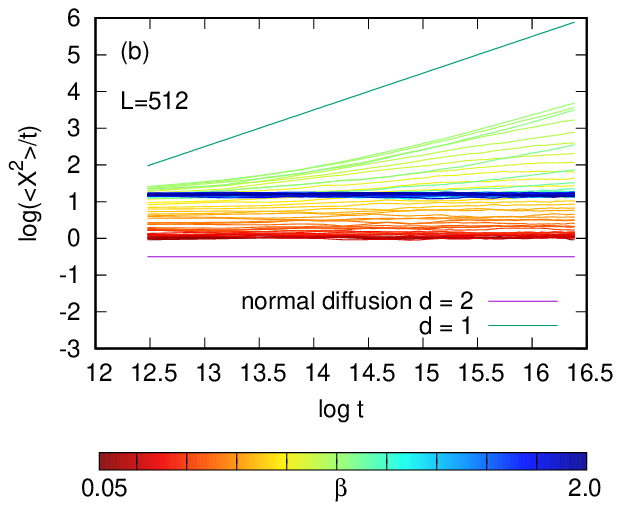}
\end{center}
\end{minipage}
\begin{minipage}{0.325\hsize}
\begin{center}
   \includegraphics[width=1\linewidth]{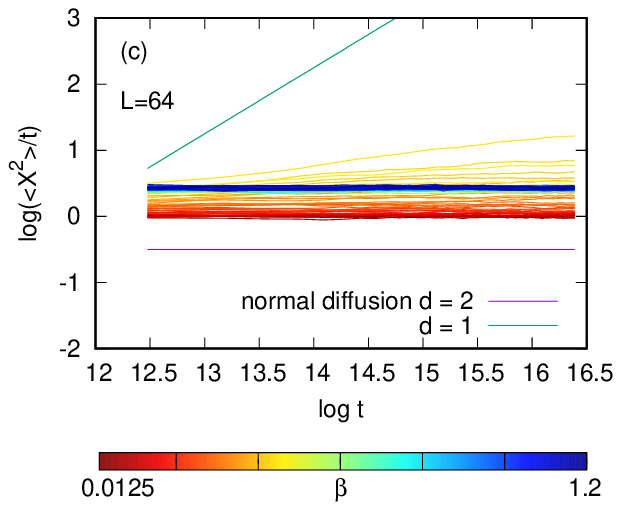}
\end{center}
\end{minipage}
\caption{Lifting variable random walk with feedback. Results for  $\log (\langle X^2 (t) \rangle/t)$ as a function of  $\log t$ for (a) 1D, (b) 2D, and (c) 3D lattices.
Each curve corresponds to MSD at each $\beta$.
The lattice sizes are (a) $L = 8192$, (b) $L = 512$, and (c) $L = 64$, respectively.}
\label{MSD_normal}
\end{figure*}

\begin{figure*}[htbp]
\begin{minipage}{0.325\hsize}
\begin{center}
  \includegraphics[width=1\linewidth]{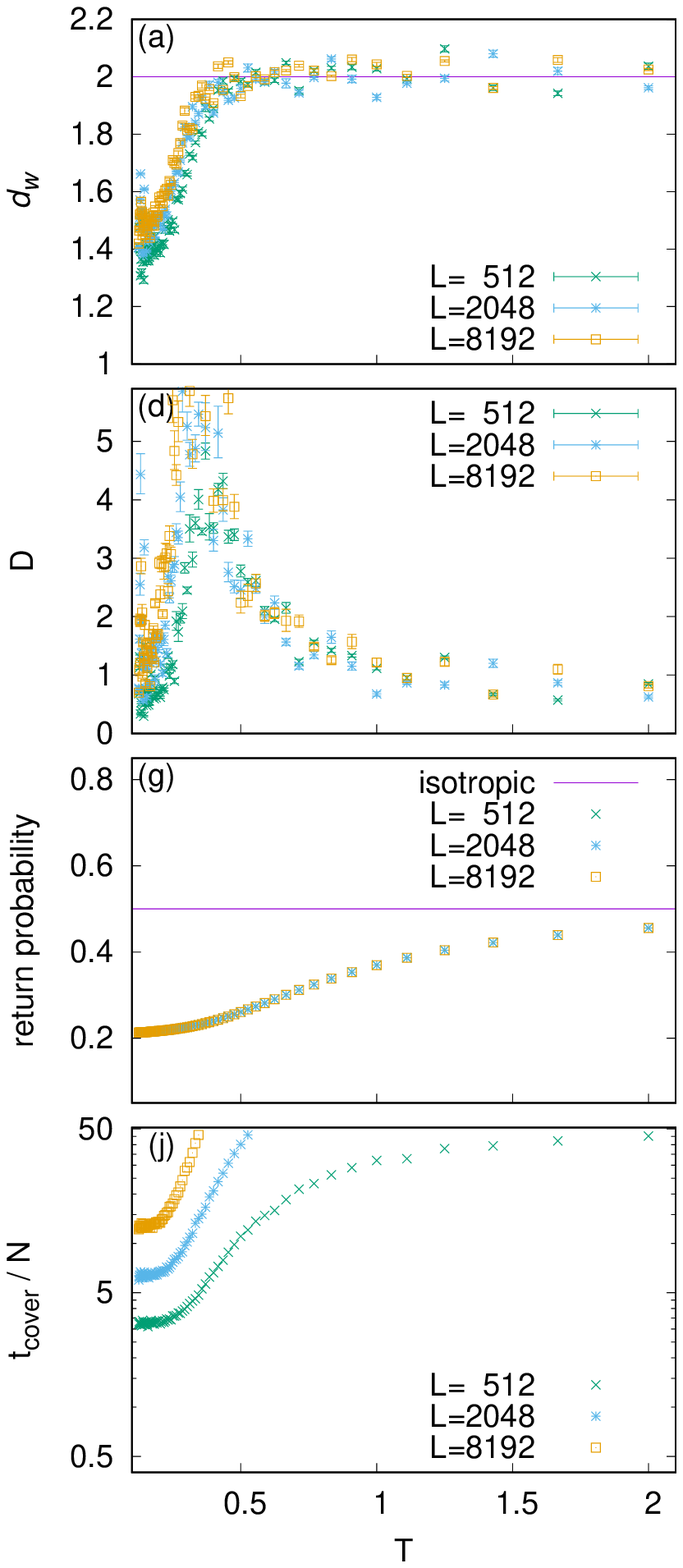}
\end{center}
\end{minipage}
\begin{minipage}{0.325\hsize}
\begin{center}
   \includegraphics[width=1\linewidth]{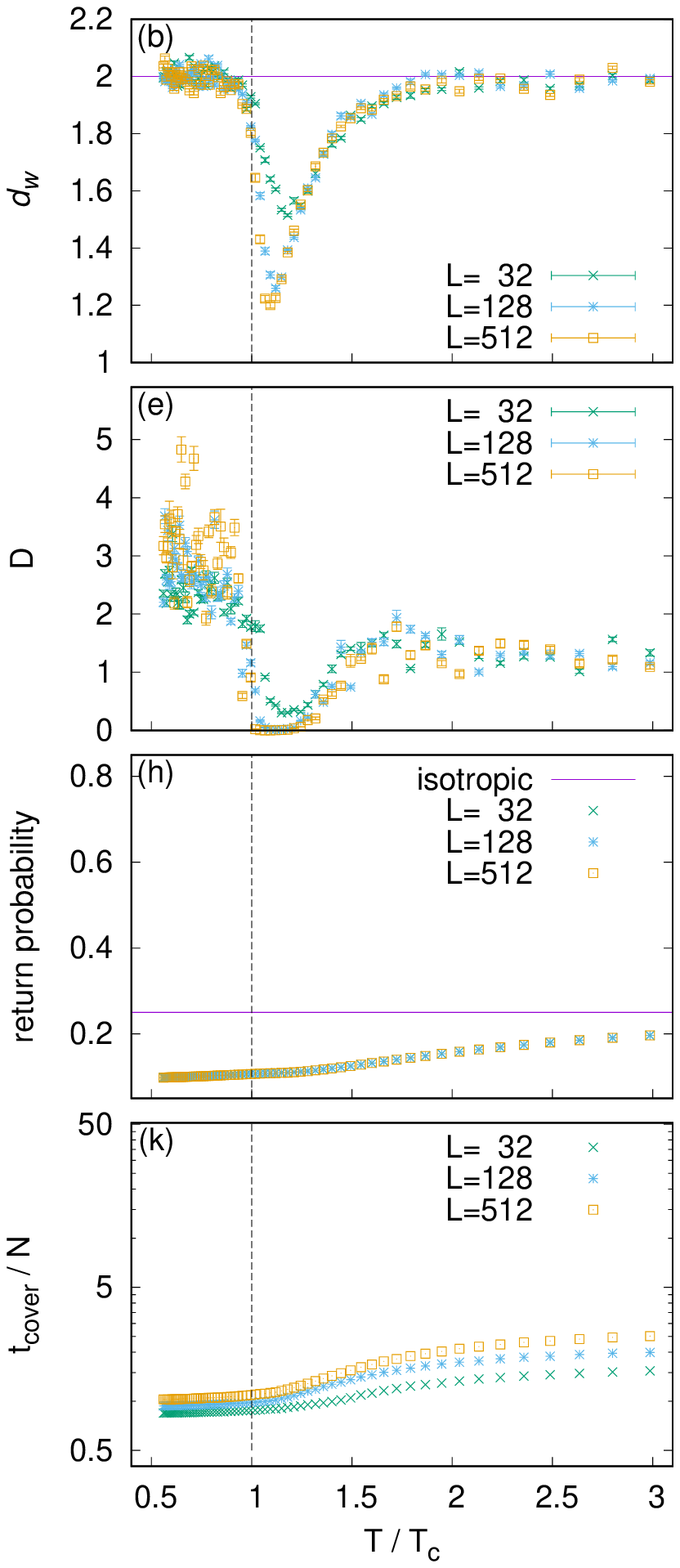}
\end{center}
\end{minipage}
\begin{minipage}{0.325\hsize}
\begin{center}
   \includegraphics[width=1\linewidth]{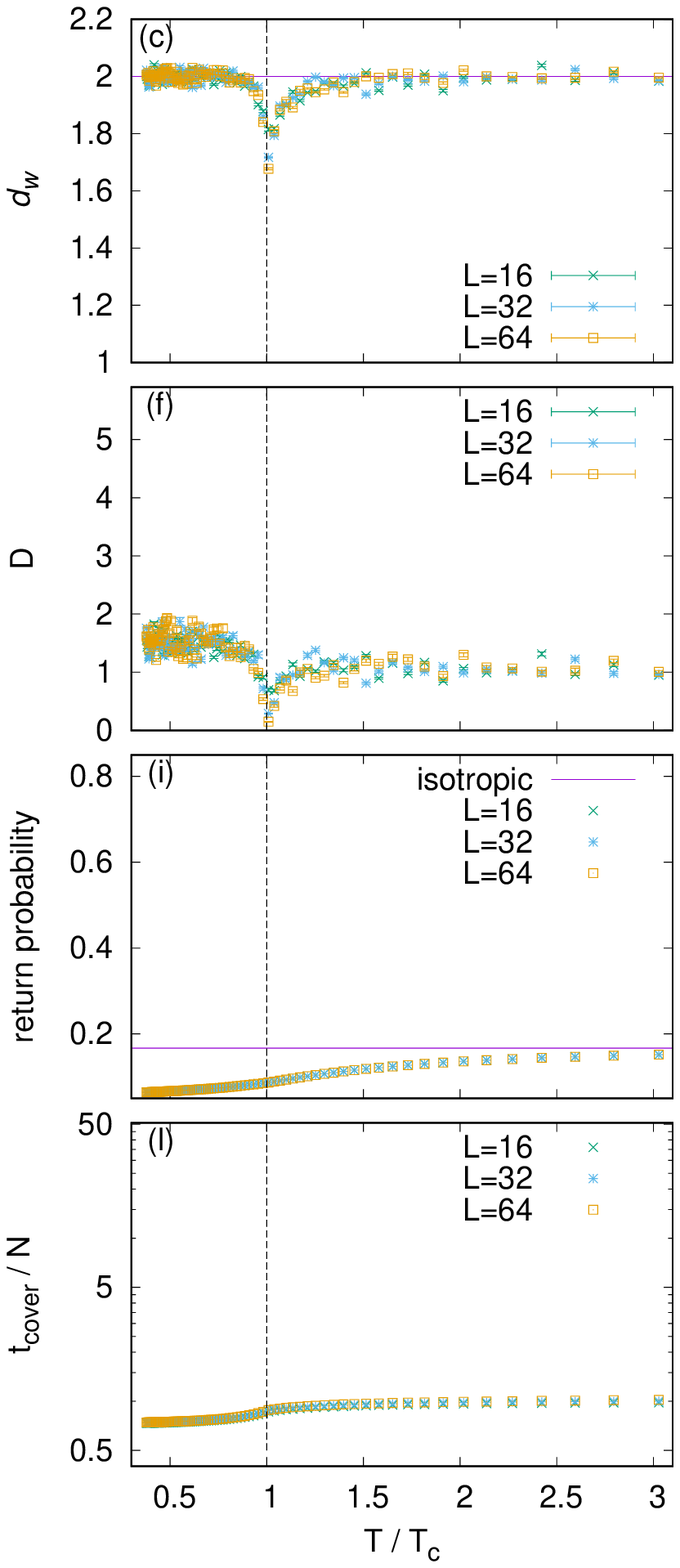}
\end{center}
\end{minipage}
\caption{\revision{Results for the random walk with feedback.
    The left, middle and right columns are results of 1D, 2D and 3D $XY$ models, respectively.
Plots of the anomalous diffusion exponent $d_\mathrm{w}$, the diffusion coefficient $D$, the return probability, and the cover time $t_\text{cover}$ are arranged from the top row to the bottom.
The dashed lines shows the critical point $T=T_\mathrm{c}$ and the purple lines in the plot of $d_\mathrm{w}$ corresponds to the normal diffusion.}}
\label{wfb}
\end{figure*}

\begin{figure*}[htbp]
\begin{minipage}{0.325\hsize}
\begin{center}
   \includegraphics[width=1\linewidth]{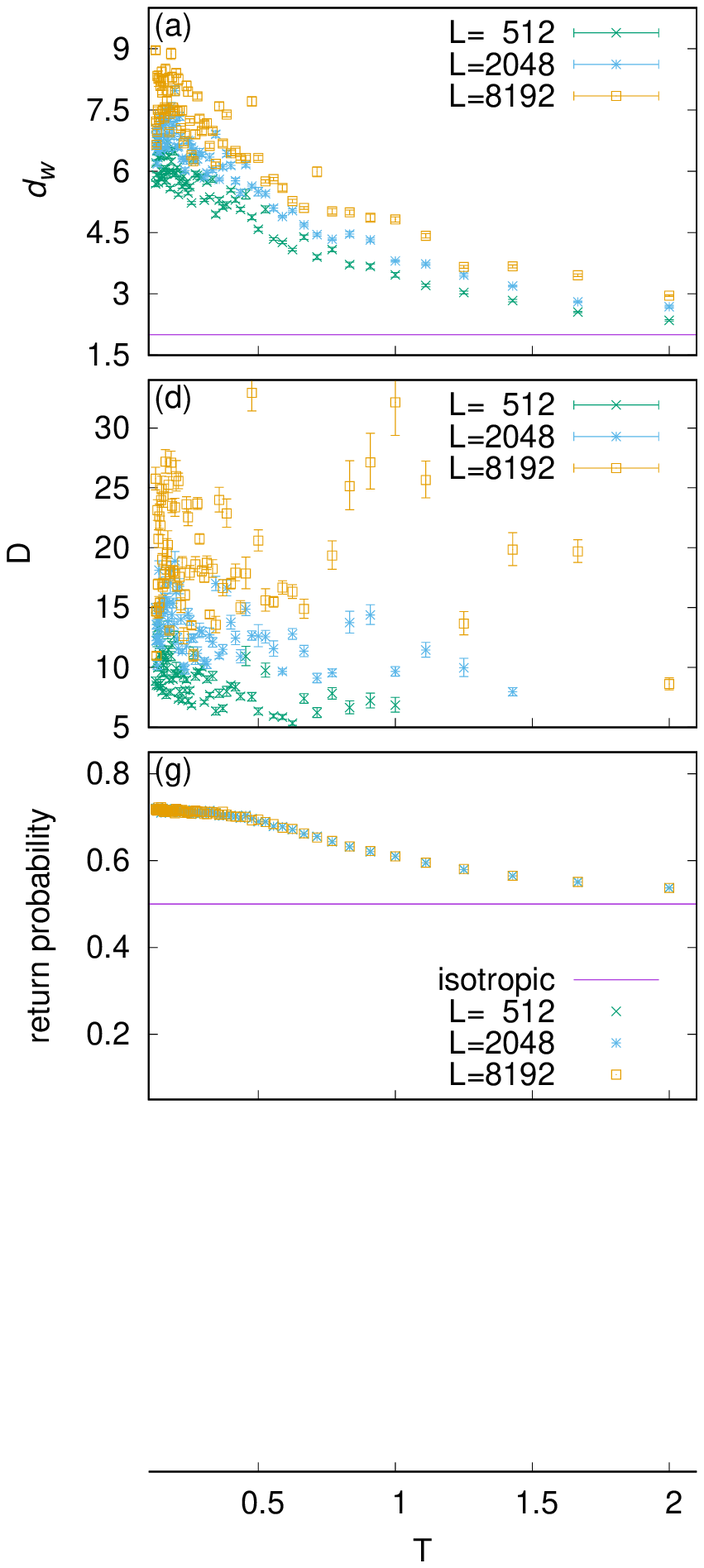}
\end{center}
\end{minipage}
\begin{minipage}{0.325\hsize}
\begin{center}
   \includegraphics[width=1\linewidth]{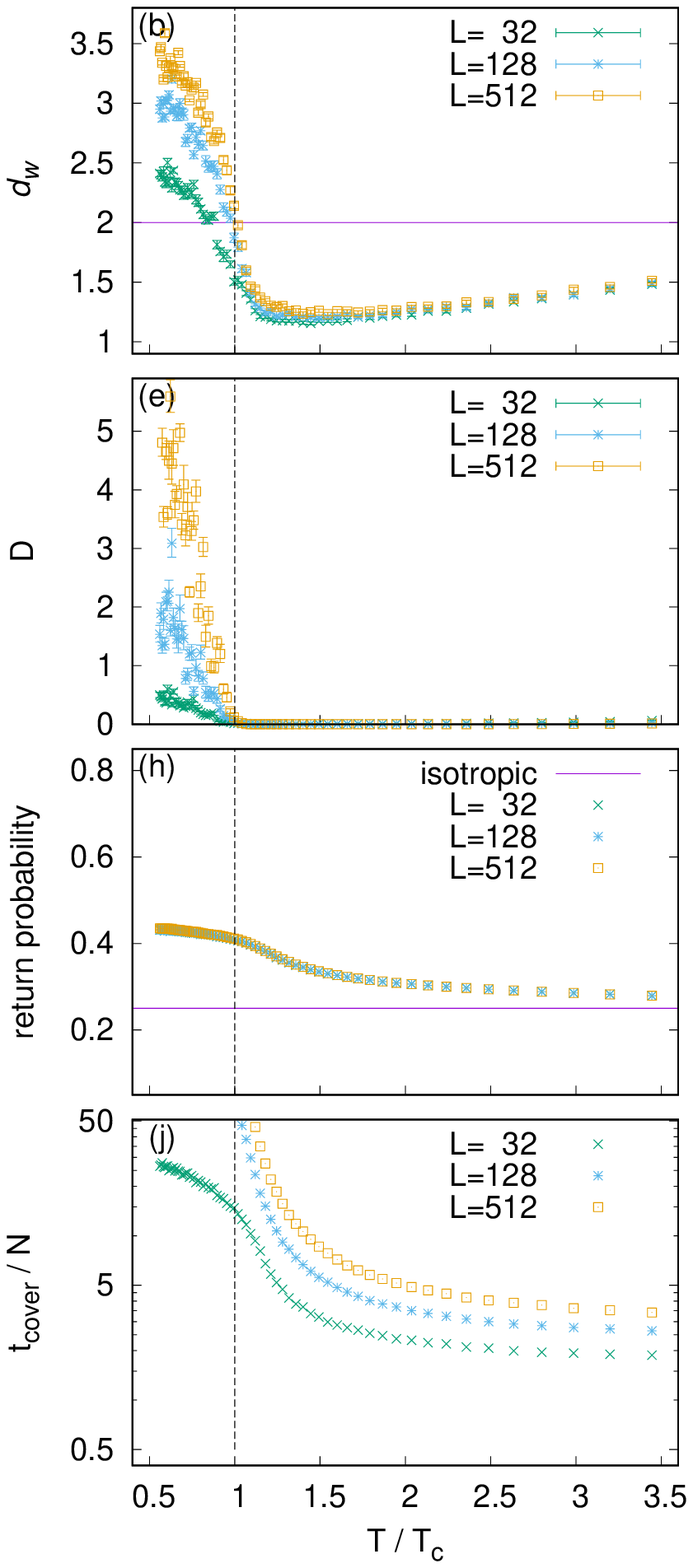}
\end{center}
\end{minipage} 
\begin{minipage}{0.325\hsize}
\begin{center}
   \includegraphics[width=1\linewidth]{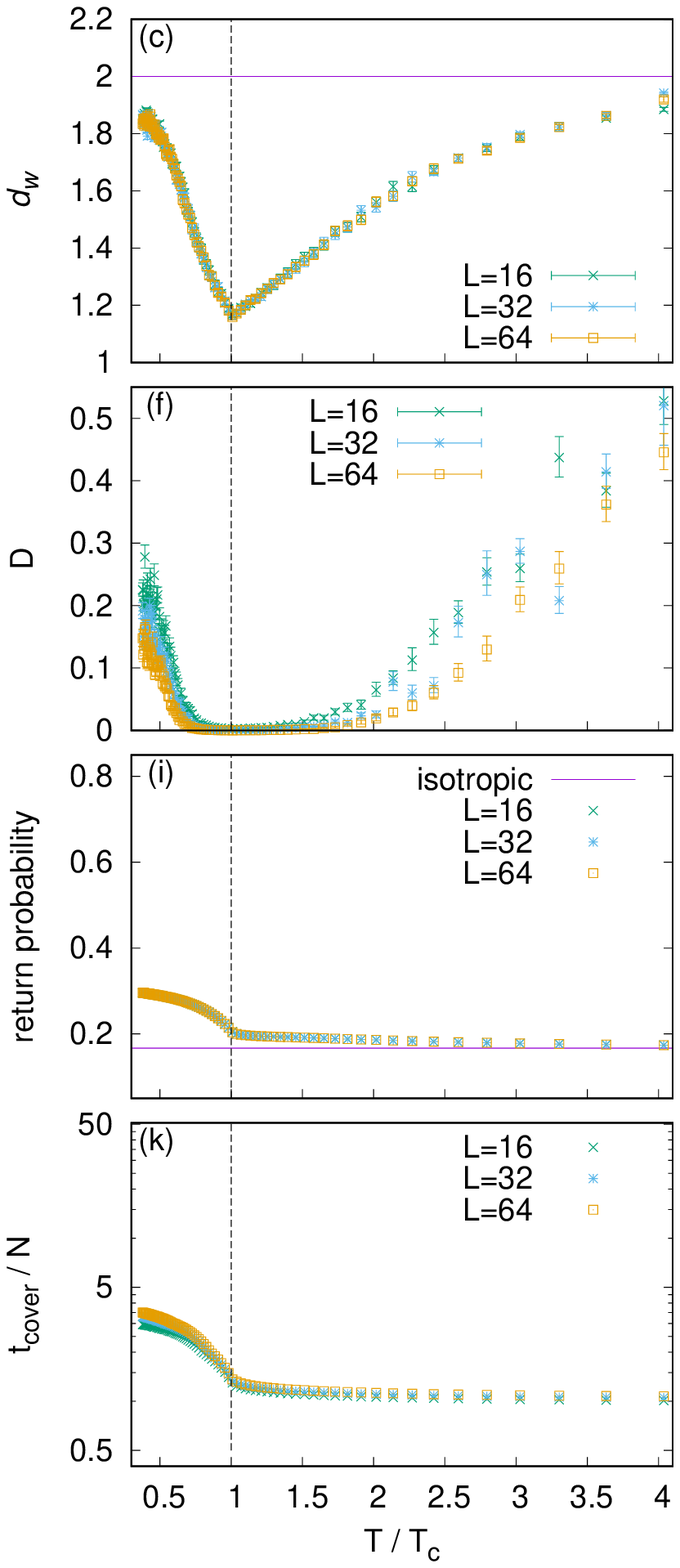}
\end{center}
\end{minipage}
\caption{\revision{Results of the random walk without feedback. Plots are arranged in the same manner as  fig.~\ref{wfb}. The cover time  $t_\text{cover}$ is undefined for 1D because the 1D walk without feedback never reaches the visit rate $1/2$ in our simulation time.}}
\label{wofb}
\end{figure*}

We analyze behaviors of the lifting variable random walk of the classical $XY$ model.
We consider the system on \revision{the} square and \revision{the} cubic lattices with periodic boundary condition with $N = L^d$ \revision{spins}, where $L$ is the length of lattice and $d=1,2,3$ is the dimension of the model.
In our results, the coupling constant $J$ is set to $1$.
The results are obtained from samples consisting of \(1000, 1000,\) and $2000$ random walks of length $50 \times N$ measured by events, for
$d=1,2$, and $3$, respectively.
We do not reset the random walk at a chain length measured by the sum of rotated angle. The initial spin configuration is equilibrated at a given inverse temperature $\beta$ before \revision{the} start.
\revision{The data are obtained in the range  $\beta_\mathrm{min}\leq\beta\leq\beta_\mathrm{max}$ with the constant interval $\Delta \beta$ where 
$(\beta_\mathrm{min},\beta_\mathrm{max},\Delta \beta)=(0.500,8.000,0.500)$, $(0.050,2.000,0.025)$, $(0.0125,1.200,0.0125)$ for $d=1,2$, and $3$, respectively.}
In fig.~\ref{MSD_normal}, we show $\log \langle X^2 (t)/t \rangle \text{-} \log t$ plots in the range of $\beta$'s. These plots clearly \revision{show} the difference of $d_\mathrm{w}$ (slope of curves).
For fig.~\ref{MSD_normal} (b) and (c), \revision{the} slope of curves increase near the critical point.
The plot indicates that the lifting variable random walk becomes \revision{superdiffusive} near the critical point.
\revision{We show the dependences of quntities of interest on the temperature $T=1/\beta$ and especially their behaviors near the critical point $T=T_\mathrm{c}$ in fig.~\ref{wfb} with feedback and in fig.~\ref{wofb} without feedback for each dimension.}

\begin{figure}
  \begin{center}
    \includegraphics[width=0.9\linewidth]{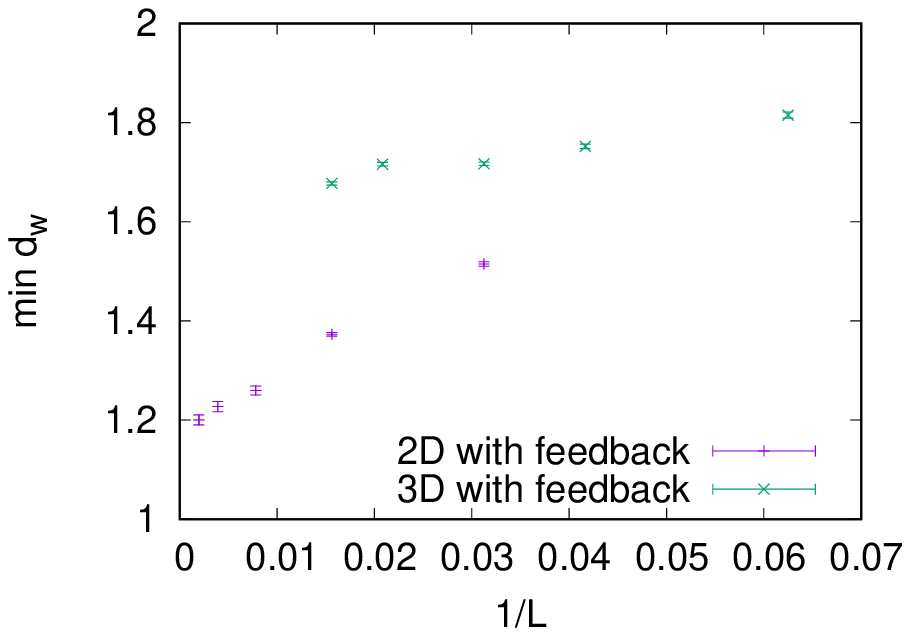}
  \end{center}
\caption{\revision{The size dependence of the minimum value $\min_T d_\mathrm{w}$ for the random walk with feedback in 2D and 3D cases. It is suggested that the values linearly converge at the infinite system size.}}\label{fsp}
\end{figure}

\subsection{Anomalous diffusion near the critical point}
To confirm the \revision{superdiffusion} \revision{observed} in fig.~\ref{MSD_normal},
we estimate $d_\mathrm{w}$ \revision{and $D$} at each \revision{$T$} by the least square method as  \revision{figs.~\ref{wfb}(a)--(f)}. 
\revision{All the cases have $d_\mathrm{w}$ larger than the ballistic movement and less than the normal diffusion.}
In the cases of 2D and 3D with feedback, the $d_\mathrm{w}$-\revision{$T$} plots have \revision{peaks near $T=T_\mathrm{c}$ and it becomes sharper with the system size increases.
The size dependence of $\min_T d_\mathrm{w}$ is shown in fig.~\ref{fsp} for sizes including those in fig.~\ref{wfb}. If the linear fit $\min_T d_{\mathrm{w}}(L)=a\times\frac1L + b$ is adopted, we obtain $\min_T d_{\mathrm{w}}(\infty)=1.20(1)$ for 2D and $1.64(1)$ for 3D.}

It may be interesting to speculate that this speedup of diffusion cancels out the critical slowdown to give the efficiency of the ECMC.
There are differences in the way peaks are formed in the cases of 2D and 3D. This can be considered to arise from the difference in nature of \revision{the} phase transition.
In the case of 1D with feedback, $d_\mathrm{w}$ \revision{has a peak at $T\rightarrow+0$} \revision{in consistent with the interpretation $T_\mathrm{c} = 0$}.

To tell whether this anomalous behavior is a consequence of the (quenched) random environment alone or the feedback in the ECMC is essential, we have performed the same measurement \revision{for the random walk without feedback defined above eq.~\eqref{msd}. Figs.~\ref{wofb}(a)--(f) show that it behaves differently from that with feedback.
Namely,} it is  \revision{subdiffusive} for 1D.
For \revision{the} 3D case, it is  \revision{superdiffusive} and $d_\mathrm{w}$ has a peak at \revision{$T=T_\mathrm{c}$}.
For \revision{the} 2D case, most surprisingly, the lifting variable random walk without feedback switches from \revision{subdiffusion} to \revision{superdiffusion} near \revision{$T=T_\mathrm{c}$}.
We also note that it has \revision{smaller} finite size effect \revision{than} \revision{the} 1D and 2D cases.

\revision{In all cases except the 1D case, the coefficient $D$ in eq.~\eqref{msd} shows a behavior similar to $d_\mathrm{w}$  while $D$ increases with the system size in the 1D case.
The difference may be due to the local nature of the 1D random walks described below.}

\subsection{Return probability and cover time}
\label{ret_cov}
We measure the return probability to have insight into the origin of the anomalous diffusion.
Here, the return probability is defined as the ratio of steps where the lifting variable \revision{has the} same value at $t$ and $t+2$ in the time series.
We obtain the return probabilities as \revision{(g)--(i) of figs.~\ref{wfb} and~\ref{wofb}}.
Interestingly, there is almost no size dependence.
The result differs from that of Markov symmetric walk which means that the lifting variable random walk has the nature of persistent random walk.

We note that also the systems near the criticality do not have \revision{a} singularity of the return probability whereas that for \revision{the} system without feedback may be singular at the critical point.

The fact that return probability with feedback at \revision{small $T$} is smaller than that of \revision{the} symmetric walk can be explained as follows. 
Imagine that the lifting variable changes its value from $k$ to a neighbor $\ell$.  The spin $k$ is rotated from $\theta_k$ to $\theta_k'=\theta_\ell+\Delta \theta$. This $\Delta \theta \in [0,\pi)$ is small at small $T$ due to eq.~\eqref{e_int}.
Then, at the next step, the spin $k$, a neighbor of $\ell$,  has less chance to be the first to meet the event among other interacting spins due to eq.~\eqref{e_int}.

The fact that return probability at \revision{small $T$} without feedback in low dimensions is larger than that of \revision{the} symmetric walk can be explained as follows. 
The lifting variable behaves following exactly the same probability every time the variable \revision{visits} a site because the spin configuration never changes. At \revision{small $T$}, the probabilistic movement becomes almost deterministic. Therefore there is a finite probability that the variable is trapped in a small region. The  simplest example of \revision{the} 1D case is a pair of sites $k, k+1$ where the variable repeats the movement $k\rightarrow k+1 \rightarrow k \rightarrow \cdots $ for a long time.

Furthermore, we investigated the visit rate which is the rate of sites visited at least once by the lifting variable.
We define `cover time' $t_\text{cover}$ as the time when the increasing visit rate reaches $1/2$.
As shown \revision{in fig.~\ref{wfb}(j)--(l) and fig.~\ref{wofb}(j),(k)}, the time heavily depends on the situation and the \revision{temperature $T$}.
As \revision{$T$} increases, $t_\text{cover}$ \revision{increases} 
in the case with feedback, while it \revision{decreases} in the case without feedback.
In \revision{the} 3D cases with or without feedback, the slope changes at the critical point almost discontinuously.
In \revision{the} 2D case without feedback \revision{on large lattices}, the time seems to diverge toward the critical point.
We consider that the feedback to spin configuration enhances diffusion of lifting variable because the difference between the results of with and without feedback \revision{suggests} that a lifting variable with feedback is not be trapped in a domain.
The behavior of return probability and cover time is consistent with the anomalous behavior of random walks.

\section{Discussions and Conclusions}
\label{sec:dis}
We have defined the lifting variable random walk in the ECMC of spin systems and have investigated the case of \revision{the} classical $XY$ model numerically.
We have shown that it becomes \revision{superdiffusion} near the critical point. This could explain \revision{the} rapid mixing and the efficiency of the ECMC even at the critical point and \revision{it} is consistent with arguments in Refs.~\cite{michel2015event, nishikawa2015event}.

Lifting variable random walk could be useful for searching a critical point of \revision{an unsolved} spin model. One could find it by just locating the parameter range where the lifting variable random walk becomes \revision{superdiffusive}.

Dynamics of the ECMC algorithm has also been investigated in Ref.~\cite{nishikawa2016event}. They have considered the joint probability distribution of the lifting variable and the spin configuration and have obtained the master equation it obeys. If the spin configuration was integrated out in their equation, we should obtain the Fokker-Plank equation that describes the lifting variable random walk with the feedback.

The present model has persistent nature. The probability of a movement depends on the previous movement. It is known that the persistent random walk becomes normal diffusion in the long time limit\cite{weiss1994aspect}. Therefore being persistent alone does not explain our result.

To confirm this observation, we run an additional simulation of \revision{the} persistent random walk, the second order Markov process, whose  position  $x(t+1)$ is equal to $x(t-1)$ with the return probability $r$ defined by the data in fig.~\ref{wfb}(g)--(i) and is equal to other possible sites at probability $(1-r)/(2d-1)$. Such walk does not indeed have the anomalous diffusion exponent $d_\mathrm{w}$ of lifting variable random walk with or without feedback. 

Super and \revision{subdiffusion} can arise in random walks with a long jump\cite{mandelbrot1982fractal, shlesinger1982random} or waiting time with a continuous distribution\cite{bouchaud1990anomalous}. It seems unlikely that the anomalous diffusion is explained in these frameworks.

\revision{subdiffusion} emerges when there are obstacles or binding sites
~\cite{SAXTON1994394, saxton1996anomalous, ellery2014characterizing}.
In our model, some spin regions could behave as obstacles,
however, our results can not be fully explained by the above theory because the walker in our system can become \revision{superdiffusive}.

In the lifting variable random walk, the current step has \revision{a} correlation with the step far before because the current spin configuration is constructed by the past movements. In this respect, the random walk in an environment with feedback is closely related to the random walk with memory\cite{schulz2000random} but has more degrees of freedom: the environment continuous spin variables. It has been reported that random walk with memory can be \revision{subdiffusive} or \revision{superdiffusive}\cite{schutz2004elephants}.

The ECMC is being developed and applied to various models including classical Heisenberg model, the $\mathrm{O}(n)$ models and others\cite{michel2017forward,nishikawa2015event,michel2017clock}. It would be interesting to investigate the lifting variable random walk of these models and to see how it is related to the underlying critical phenomena. This is left for future works.

\acknowledgments
We are grateful to Shinji Iida and Junta Matsukidaira for discussions.
We also thank an anonymous referee for bringing our attention to the anomalous diffusion exponent for inifinite size system.

\bibliographystyle{eplbib} 
\bibliography{mc,ecmc,rw,xy}

\end{document}